\input harvmac
\newcount\figno
\figno=0
\def\fig#1#2#3{
\par\begingroup\parindent=0pt\leftskip=1cm\rightskip=1cm
\parindent=0pt
\baselineskip=11pt
\global\advance\figno by 1
\midinsert
\epsfxsize=#3
\centerline{\epsfbox{#2}}
\vskip 12pt
{\bf Fig. \the\figno:} #1\par
\endinsert\endgroup\par
}
\def\figlabel#1{\xdef#1{\the\figno}}
\def\encadremath#1{\vbox{\hrule\hbox{\vrule\kern8pt\vbox{\kern8pt
\hbox{$\displaystyle #1$}\kern8pt}
\kern8pt\vrule}\hrule}}

\overfullrule=0pt

\Title{UK-03-04, Brown-HET-1350}
{\vbox{\centerline{Large-N Collective Fields and Holography}}}
\smallskip
\centerline{Sumit R. Das}
\smallskip
\centerline{\it Department of Physics and Astronomy}
\centerline{\it University of Kentucky, Lexington, KY 40506.}
\centerline{\tt das@pa.uky.edu}
\smallskip
\centerline{and}
\smallskip
\centerline{Antal Jevicki}
\smallskip
\centerline{\it Department of Physics}
\centerline{\it Brown University, Providence, RI 02192}
\centerline{\tt antal@het.brown.edu}
\bigskip

\medskip

\noindent

We propose that the euclidean bilocal collective field theory of
critical large-N vector models provides a complete definition of the
proposed dual theory of higher spin fields in anti de-Sitter
spaces. We show how this bilocal field can be decomposed into an
infinite number of even spin fields in one more dimension. The
collective field has a nontrivial classical solution which leads to a
$O(N)$ thermodynamic entropy characteristic
of the lower dimensional theory, as required by general
considerations of holography. A subtle cancellation of the entropy
coming from the bulk fields in one higher dimension with $O(1)$
contributions from the classical solution ensures that the subleading
terms in thermodynamic quantities are of the expected form.  While the
spin components of the collective field transform properly under
dilatational, translational and rotational isometries of $AdS$,
special conformal transformations mix fields of different spins
indicating a need for a nonlocal map between the two sets of fields.
We discuss the nature of the propagating degrees of freedom through a
hamiltonian form of collective field theory and argue that nonsinglet
states which are present in an euclidean version are related to
nontrivial backgrounds.

\def\half{{1\over 2}}

\def\vphi{{\vec \phi}}
\def\vx{{\vec x}}
\def\vy{{\vec y}}
\def\vu{{\vec u}}
\def\vv{{\vec v}}
\def\vm{{\vec m}}
\def\vn{{\vec n}}
\def\vk{{\vec k}}
\def\cS{{\cal S}}
\def\cD{{\cal D}}
\def\tA{{\tilde A}}
\def\tsigma{{\tilde \sigma}}
\def\teta{{\tilde \eta}}
\def\sN{{\sqrt{N}}}
\def\vp{{\vec p}}
\def\tzeta{{\tilde \zeta}}

\def\tx{{\tilde x}}
\def\ty{{\tilde y}}
\def\hv{{\hat v}}
\def\tu{{\tilde u}}
\def\tv{{\tilde v}}

\newsec{Introduction and summary}

It has been known for a long time that theories of $N \times N$
matrices, e.g. gauge theories, become string theories at large $N$, with
${1\over N}$ playing the role of string coupling constant
\ref\thooft{G. 't Hooft, {\it Nucl. Phys.} {\bf B72} (1974) 461.}. 
A new element of this correspondence was learnt
in the early 90's, viz. the string theory lives in a higher dimensional
space. 
An early example is the quantum mechanics of a single hermitian matrix
$M_{ij}(t)$ - the $c=1$ matrix model.  In this case the string theory
is in $1+1$ dimensions, whose only dynamical field
is a massless scalar. This scalar is related to the density of
eigenvalues $\phi (x,t)$ of the matrix, so that the space of
eigenvalues $x$ provides the extra coordinate \ref\djev{S.R. Das and
A. Jevicki, {\it Mod. Phys. Lett.} {\bf A5} 1060.}. The
hamiltonian for this collective field can be written down
following \ref\jevsaka{A. Jevicki and B. Sakita,
{\it Nucl. Phys.} {\bf B 165} (1980) 511.}. 
The classical value of $\phi (x,t)$ corresponds to the linear dilaton
background of string theory while the fluctuations are related to the
massless scalar \foot{The detailed relationship
between $\phi(x,t)$ and the scalar which couples simply to the
worldsheet is, however, rather complicated
\ref\ceqone{See e.g. J. Polchinski, {\tt hep-th/9411028};
A. Jevicki, M. Li and T. Yoneya, {\it Nucl.Phys.} {\bf B448} (1995)
{\tt hep-th/9504091}
}}. Thus the singlet sector of the model
contains the propagating 
degrees of freedom of the two dimensional string theory.
This is, in a sense, holography \ref\holo{G. 't Hooft, 
in ``{\it Salamfest}'' (1993) 0284,
{\tt gr-qc/9310026}; L. Susskind, 
{\it J. Math. Phys.} {\bf 36} (1995) 6377, {\tt hep-th/9409089}}. 
However, the singlet sector thermodynamic entropy of the
collective field theory 
is $O(1)$ rather than $O(N^2)$ and is 
that of a $1+1$ dimensional theory rather than that of
a $0+1$ dimensional theory \djev. 
This is different from what one expects from
the holographic principle \holo. In fact the contribution of
non-singlet states should be significant in the thermodynamic
partition function at high temperatures which is precisely the regime
where a $0+1$ dimensional behavior with entropy proportional to $N^2$
is expected. Such non-singlet states are, however, not part of the
perturbative string theory spectrum.

The most concrete example of holography is of course the $AdS/CFT$
correspondence \ref\malda{J. Maldacena, {\it Adv. Theo. Math. Phys.} 
{\bf 2}
(1998) 231, {\tt hep-th/9711200}}
\ref\gkpw{S. Gubser, I. Klebanov and
       A. Polyakov, {\it Phys. Lett.} {\bf B428} (1998) 105, {\tt
       hep-th/ 9802109}; E. Witten, {\it Adv. Theo. Math. Phys.} {\bf
       2} (1998) 253, {\tt hep-th/9802150}.}. 
Here, the high temperature thermodynamics is dominated
by $AdS$ black holes
\ref\susswit{L. Susskind and E. Witten, 
{\tt hep-th/9805114}}. For $AdS_{d+1}$ with length scale $R$, and a
$d+1$-dimensional Newton constant $G$, the entropy of a black hole
$S_{bulk}$ is related to the temperature $T$ by the relation
\eqn\one{S_{bulk}~G \sim (T~R^2)^{d-1}}
On the other hand, the conformal field theory on the boundary at the
same temperature should have an entropy $S_{boundary}$ given by
\eqn\two{S_{boundary} \sim N_f (TR)^{d-1}}
where $N_f$ is the number of degrees of freedom.
These two expressions agree if
\eqn\three{ {G \over R^{d-1}} \sim {1 \over N_f}}
- a relation which is satisfied for all known examples of 
$AdS/CFT$.

There are black holes in $1+1$ dimensional string theory as well
\ref\msw{G. Mandal, A. Sengupta and S.R. Wadia,
{\it Mod.Phys.Lett.} {\bf A6} (1991) 1685.; E. Witten, 
{\it Phys. Rev.} {\bf D44} 314}. However despite
considerable effort \ref\twodbh{S.R. Das, {\it Mod. Phys. Lett.}
{\bf A8} (1993) 69; S.R. Das, {\it Mod. Phys. Lett.}
{\bf A8} (1993) 1331; A. Dhar, G. Mandal and S.R. Wadia,
{\it Mod. Phys. Lett.}
{\bf A7} (1992) 3703;
A. Dhar, G. Mandal and S.R. Wadia,
{\it Mod. Phys. Lett.}
{\bf A8} (1993) 1701; A. Jevicki and T. Yoneya,
{\it Nucl. Phys.} {\bf B 411} (1994) 64.}, such
black holes are not fully understood in the matrix model, though
there has been significant recent progress \ref\kkk{V. Kazakov,
D. Kutasov and I. Kostov, {\it Nucl.Phys.} 
{\bf B622} (2002) 141, {\tt hep-th/0101011};
S. Alexandrov, V. Kazakov and I. Kostov, {\tt hep-th/0302106}}. 
The above discussion shows that these black
holes are intimately related to non-singlet states, along the
lines of \kkk. In the $AdS/CFT$ examples all the physical states
are singlets in any case, which may be the reason why the thermodynamics
is reproduced faithfully.

In models with matrices, there are an exponentially growing number of
single trace singlet operators, which is one reason why they represent
string theories. In this paper we will consider models with fields in
the vector representation of groups like $O(N)$ or $U(N)$. These
models are known to be solvable in the large-$N$ limit
\ref\vector{See e.g. S. Coleman ``1/N'' in 
Erice lectures, 1979 and references to the original literature therein}. 
The singlet operators can involve products of
pairs of fields and therefore there are no exponentially growing number of
``single particle'' states. Indeed, the Feynman diagrams are made of
bubbles and resemble branched polymers rather than string worldsheets.

In a recent work, Klebanov and Polyakov \ref\kp{I. Klebanov and
A.M. Polyakov, {\tt hep-th/0210114}} 
have proposed that critical vector models are dual to certain
higher spin gauge theories \ref\higher{C. Fronsdal,
{\it Phys. Rev.} {\bf D18} (1978) 3624; E. Fradkin and
M. Vasiliev, {\it Phys. Lett.} {\bf B 189} (1987) 89;
E. Fradkin and
M. Vasiliev, {\it Nucl. Phys.} {\bf B291} (1987) 141;
M. Vasiliev, {\tt hep-th/9910096}}
\ref\higherb{E. Sezgin and P. Sundell,
{\tt hep-th/0205132}; J. Engquist, E. Sezgin and
P. Sundell, {\tt hep-th/0207101}; M. Vasiliev,
{\it Int. J. Mod. Phys.} {\bf D5} (1996) 763} 
defined on $AdS$
spaces. Such higher spin fields include gravity, but are not
string theories. (Higher spin gauge theory
with $N=8$ supersymmetry in $AdS_5$ is, however, related to 
{\it free} $N=4$ Yang-Mills theory and hence to $\alpha ' \rightarrow
\infty$ limit of IIB string theory \ref\higherc{P. Haggi-Mani and
B. Sundborg, {\tt hep-th/0002189}; B. Sundborg,
{\tt hep-th/0103247}; E. Witten, {\tt http://theory.caltech.edu/
jhs60/witten/1.html}; E. Sezgin and P. Sundell, {\tt hep-th/0105001}}
\ref\higherd{
A. Mikhailov, {\tt hep-th/0201019}}). The complete set of interaction
in these theories is still not known.

Of particular interest is the critical three (euclidean)
dimensional $O(N)$ vector model
\eqn\four{S = \int d^3 x [ (\partial \vphi)\cdot (\partial \vphi)
+ {\lambda \over 2 N}(\vphi \cdot \vphi)^2 ]}
This has two fixed points. In terms of a running coupling $\lambda
(k)$ these are at $\lambda (k) = 0$ and $\lambda (k) = 4k$. According
to the proposal of \kp, the three dimensional conformal field theories
at these fixed point are dual to a theory of higher spin fields with
one field for each even spin defined on $AdS_4$ in two different
senses.  The correspondence of bulk and boundary quantities for the
theory at the nontrivial fixed point is standard, with the generating
functional for correlators of singlet operators being equal to the
effective action in the bulk with the boundary values of the fields
set equal to the currents . For the theory at the gaussian fixed point
the correspondence proceeds via a legendre transform of the generating
functional as in \ref\kw{I. Klebanov and E. Witten,
{\it Nucl. Phys.} {\bf B556} (1999) 89, {\tt hep-th/9905104}.}

In this paper we show that standard methods of collective field theory
can be used to start with a vector model in euclidean space and 
{\it construct} a theory of
even spins in one higher dimension.  The main ingredient is the fact
that all singlet correlations of the model may be expressed in terms
of a bilocal field $\sigma (\vx, \vy)$ 
\ref\cjt{J.Cornwall,R.Jackiw and E.Tomboulis,
{\it Phys.Rev.} {\bf D10} (1974) 2428.} ,\ref\jevsak{A. Jevicki
and B. Sakita, {\it Nucl.Phys.} {\bf B185} (1981) 89.} 
\eqn\five{\sigma (\vx, \vy) = {1\over N} \vphi (\vx) \cdot
\vphi (\vy)}
The higher dimensional space is made out of the center of mass coordinate
\eqn\six{ \vu = \half (\vx + \vy)}
and the magnitude $r$ of the relative coordinate
\eqn\seven{ \vv = \half (\vx - \vy)~~~~~~~~~r^2 = \vv \cdot \vv}
A spherical harmonic decomposition in the angles of the coordinate
$\vv$ then yields a collection of higher spin fields in one higher
dimension. The symmetry of the collective field $\sigma (\vx,\vy)$
under interchanges of $\vx$ and $\vy$ implies that these spins are
zero or even integers. While this construction works for any number
of dimensions and for generic values of the coupling, things become
more interesting when the theory is conformally invariant. By
examining the transformation properties of the field $\sigma (\vx,\vy)$
under conformal transformations we find that the various spin
components of $\sigma (u,v)$ transform as tensors under translation,
rotation and dilatation isometries of (euclidean) $AdS$ space, where 
$r$ is a coordinate in the following form of the $AdS$ metric :
\eqn\eight{ds^2 = {1\over r^2}[dr^2 + d\vu \cdot d\vu]}
However these components {\it do not} transform properly under special
conformal isometries.
We suspect that this indicates
that these modes are related to the standard fields in
$AdS$ by a field redefiniton which is possibly nonlocal. This is
similar to the fact that in the $c=1$ matrix model the collective
field is not really the field which follows from a worldsheet formulation
of the dual string theory. Our considerations may be easily extended
to vector models with other symmetries, in which case one would get
even spins as well.

From the point of view of the vector model, the bilocal collective
field represents a collection of local composite operators. This may
be seen by performing a Taylor expansion in the coordinate $\vv$ so
that one has
\eqn\sevenc{\sigma (\vx,\vy) = \vphi (\vx) \vphi (\vx) + [\vphi (\vx)
\partial_i \partial_j \vphi (\vx) - (\partial_i \vphi
(\vx))(\partial_j \vphi (\vx))]v^i v^j + \cdots}
The coefficients of powers of $v$ in this
expansion are related, but {\it not identical}, 
to the infinite set of conserved currents of the
free theory which are conjectured to be the operators dual to the
higher spin fields. The nontrivial relationship between these currents
and the coefficients in the expansion \sevenc\ is another indication
of the fact that the relationship between $\sigma$ and the higher spin
fields is nonlocal. Nevertheless, since these currents are all
contained in the bilocal field one can construct bulk fields in $AdS$
by folding in with the appropriate bulk-boundary Green's function
as in \higherd.

In this paper we construct the euclidean collective field theory with
special attention to subleading effects in ${1\over N}$. As is well
known, the collective field has a nontrivial classical solution. In our
interpretation this 
provides the four dimensional 
``spacetime'' on which the physical excitations propagate.
By considering fluctuations around the classical solution we
demonstrate the existence of a nontrivial IR stable fixed point in
three dimensions and reproduce the well known results for conformal
dimensions of composite operators at this fixed point.

One finite $N$ effect which is not discussed in this
paper in detail is the emergence of an {\it exclusion
principle}. Consider the problem on a finite lattice with $M$ sites in
each of the $d$ directions. Then the functional integral over $\phi^i$
is a multiple integral over $NM^d$ variables. However the functional
integral over the collective field
$\sigma(\vx,\vy)$ is a multiple integral over $M^{2d}$ variables. Thus
if $N < M^d$ there are too many degrees of freedom in the collective
field. This would lead to rather nontrivial constraints on $\sigma$.
In terms of fourier modes of $\sigma$ this means that all these modes
are not independent. Roughly speaking, for each component of the
momenta, one may regard the first $N^{1/d}$ values to be independent,
while the remaining modes are related to these by nontrivial
relations. For the bulk theory this is a kind of exclusion principle 
which has appeared in
the context of $AdS/CFT$ correspondence and which arises because of
the same reason \ref\exclusion{J. Maldacena and A. Strominger,
{\it JHEP} {\bf 9812} (1998) 005, {\tt hep-th/9804085}; A. Jevicki
and S. Ramgoolam, {\it JHEP} {\bf 9904} (1999) 032, {\tt hep-th/9902059}}.

One of our main results relates to the thermodynamics of the model.
From the point of view of the critical vector model defined in $d$
euclidean dimensions, the finite temperature properties are those
appropriate to $d$ dimensions. Furthermore, one would expect that the
leading free energy and the entropy are both proportional to $N$.  This,
however, appears quite mysterious from the point of view of a $d+1$
dimensional theory where ${1\over {\sqrt N}}$ appears as a {\it
coupling constant}, so that the natural expectation is that in a
${1\over N}$ expansion the leading entropy comes from the free theory
and hence of $O(1)$.  We show that the leading thermodynamic behavior
is a {\it classical} contribution in the collective field
theory coming from the presence of a nontrivial classical solution. 
Perhaps more significantly, there is a partial cancellation
between the $O(1)$ contribution obtained by integrating out
fluctuations and a $O(1)$ term present in the classical action. In
particular, for the gaussian fixed point this cancellation is complete
and ensures that the entire answer is $O(N)$ as expected. We speculate
that this result is indicative of the presence of black holes in the
bulk theory. The fact that the thermodynamics is reproduced correctly
is an indication that these models, like string theory examples, have
the right ingredients to provide a holographic description of theories
containing gravity.  Unlike string theory examples, however, we have
an explicit construction of the higher dimensional theory in terms of
the fields of the lower dimensional theory. The hope is that this will
facilitate a better understanding of holography.

This explicit construction in fact shows a special feature of the bulk
theory. We show that the interactions of the collective field theory
have a coupling constant ${1\over {\sqrt{N}}}$ {\it with no other free
parameter}. On the other hand the bulk theory has {\it a priori} two
dimensional parameters, the Newton constant $G$ and the $AdS$ scale
$R$. The collective field theory seems to indicate, however, that the
bulk theory must be characterized by only the dimensionless
combination ${G \over R^2}$. This should be exactly ${1\over
N}$. Indeed, this is exactly what is required in $d=3$ from \three.
This is related to the fact that the conformal field theory of the
vector model lives at {\it fixed points} rather than on {\it fixed
lines} there is no free coupling constant which would be the analog of
the gauge coupling constant for string theory on $AdS_5 \times
S^5$. In this sense this is similar to M-theory examples of
holography.

The {\it euclidean} collective field however, contain {\it more} than
{\it propagating} degrees of freedom. This is related to the fact that
the collective field is made out of currents of the vector model
which are conserved. Thus all the normal modes of the current do not
create independent and orthogonal states. This feature is in fact well
known in the $AdS/CFT$ correspondence. For example the operator in the
dual theory which represents the bulk graviton is the energy momentum
tensor which seems to have more components than the graviton. However
the energy momentum tensor is conserved, which reduces the number of
independent modes to the correct value.
  
To formulate the theory in terms of the physical propagating degrees
of freedom it is more useful to consider a hamiltonian formulation of
collective field theory where the dynamical variables are $\psi
(\tx,\ty,t)$ where $\tx$ and $\ty$ denote points in {\it space} and
$t$ is the time.
These have canonical conjugate momenta
$\Pi(\tx,\ty, t)$ and one can derive a hamiltonian which reproduces the
correlators of such singlet operators. However, like other examples of
hamiltonian collective fields (notably the $c=1$ matrix model), it is
difficult to describe nontrivial {\it backgrounds} and to describe the
finite temperature thermodynamics fully.

In section 2 we derive the euclidean collective field action for any
dimension after a careful derivation of the jacobian, 
discuss the saddle point solution at large $N$, the action for quadratic
fluctuations and the $O(1)$ partition function, the nature of
interactions and the appearance of the nontrivial fixed point in
$d=3$. 
In section 3 we discuss the
implications of our results for the holographic correspondence : the
finite temperature thermodynamics, the identification of dilatation,
rotation and translation isometries of the bulk, the nature of
interactions in the bulk theory and the question of physical
propagating modes and its relationship to hamiltonian collective field
theory. Section 4 contains conclusions and comments.

\newsec{Euclidean Collective Field Theory : Saddle point solution and
fluctuations }

We start with the following action in $d$ euclidean dimensions
\eqn\nine{S[\vphi] = \int d^d x [ (\partial \vphi)\cdot (\partial \vphi)
+ m^2 \vphi \cdot \vphi + 
{\lambda \over 2 N}(\vphi \cdot \vphi)^2 ]}
The collective field is defined as in \five. The collective field
action $\cS[\sigma]$ is then defined via the relation
\eqn\ten{ \int \cD \vphi (x)~e^{-S[\vphi]}
= \int \cD \sigma (\vx,\vy)~e^{-\cS[\sigma (\vx,\vy)]}}

\subsec{Derivation of the action}
The action $\cS$ has a piece $\cS_0$ which comes from rewriting \nine\ in 
terms of $\sigma$ and a second piece coming from the Jacobian $J$ in
the change of variables in the path integral measure.
\eqn\eleven{ \cS = \cS_0 - \log~J}
$\cS_0$ may be written down easily from \nine\
\eqn\twelve{\cS_0 = N
 \int d^d x \{ -[\nabla_x^2 \sigma(\vx,\vy)]_{\vx=\vy}
+ m^2 \sigma (\vx,\vx) + {\lambda \over 2 } (\sigma(\vx,\vx))^2 \} }

To treat singular terms which appear in $J$ we will work on a 
square lattice with $M$ points in each direction. The fields $\phi$
will be also rescaled appropriately to make them dimensionless, so
that the lattice spacing disappears from the expressions. At the
end of the calculation one may of course restore the lattice spacing
and take the continuum and thermodynamic limit. A point on the
lattice will be denoted by an integer-valued vector $\vm$. The
Jacobian can be then determined by comparing Dyson Schwinger
equations for invariant correlators obtained from the ensembles on
the two sides of \ten\ \jevsak, \ref\wadia{S.R. Wadia,
{\it Phys. Rev.} {\bf D 24} (1981) 970} \ref\mello{R. de Mello Koch and J.P.
Rodrigues, {\it Phys.Rev.} {\bf D54} (1996) 7794,
{\tt hep-th/9605079}}. First consider the identity
\eqn\thirteen{\int [\cD \phi] {\delta \over \delta \phi^i (\vm)}
(\phi^i (\vm') F[\sigma])~e^{-S[\vphi]} = 0}
for some arbitrary functional $F[\sigma]$ 
of the bilocal collective field. This
leads to the equation
\eqn\fourteen{<N \delta_{\vm,\vm'}~F[\sigma]>
+< \phi^i(\vm') {\delta F \over \delta \phi^i(\vm)}>
- <\phi^i (\vm ') {\delta S \over \delta \phi^i (\vm)}~F[\sigma]> = 0}
where the averages are evlauated with the action $S[\vphi]$. Using
\eqn\fourteena{{\delta \over \delta \phi^i (\vm)}
= \sum_{\vm_1,\vm_2}{\delta \sigma(\vm_1,\vm_2) \over \delta \phi^i
(\vm)}
{\delta \over \delta \sigma (\vm_1,\vm_2)} =
\sum_{\vm_1}\phi^i(\vm_1)[{\delta \over \delta \sigma (\vm_1,\vm)}
+ {\delta \over \delta \sigma (\vm,\vm_1)}]}
\fourteen\ becomes
\eqn\fourteenb{N \delta_{\vm,\vm'} <F> + 2\sum_{\vm_1}\sigma
(\vm_1,\vm')[{\delta F \over \delta \sigma (\vm_1,\vm)}
- F {\delta S \over \delta \sigma (\vm_1,\vm)}] = 0}

Next 
consider a change of variables to the collective fields 
$\sigma(\vm,\vm')$ and consider the identity
\eqn\fifteen{ \sum_{\vm_1} \int [\cD \sigma]~{\delta \over \delta 
\sigma (\vm_1, \vm)} ( \sigma (\vm_1, \vm')~J[\sigma]~F[\sigma]
~e^{-\cS}) = 0}
where $J[\sigma]$ is the jacobian which we want to determine. 
The averages $\ll \cdot \gg$ are defined in the ensemble with the
action $\cS$ and measure $[\cD \sigma] J[\sigma]$. Note that for
any observable $A$ one has the identity
\eqn\sixteen{ <A> = \ll J~A \gg} 
Then \fifteen\ becomes
\eqn\seventeen{\eqalign{ 
M^d \delta_{\vm,\vm'} \ll~F~\gg + &
\sum_{\vm_1} \ll~\sigma(\vm_1, \vm')~{\delta~\log~J \over
\delta \sigma(\vm_1,\vm)}~F~\gg 
 + \sum_{\vm_1} \ll~\sigma(\vm_1, \vm')~{\delta F \over
\delta \sigma(\vm_1, \vm)}~\gg \cr
& - \sum_{\vm_1} \ll~\sigma(\vm_1, \vm')~F~
{\delta \cS \over \delta \sigma (\vm_1,\vm)}~\gg = 0}}
Since $F$ is arbitrary, comparing \seventeen\ with \fourteenb\ one gets
an equation for $J$
\eqn\eighteen{\sum_{\vm_1} \sigma (\vm_1, \vm') {\delta~\log~J \over
\delta \sigma(\vm_1,\vm)} = \half (N - 2 M^d) \delta_{\vm,\vm'}}
which may be solved by
\eqn\nineteen{ \log~J =\half (N - 2 M^d)~{\rm Tr} \log~\sigma}
upto a constant.
Here the trace is taken over the indices $\vm$.
 The final expression \nineteen\ can be of course written
in continuum notation, in which the role of the factor $M^d$ is
given by $V \delta^d (0)$ where $V$ is the volume of the $d$
dimensional space. 

\subsec{The saddle point solution}

Since both terms in \eleven\ have pieces proportional to $N$,
the functional integral may be performed by a saddle point
method at $N \rightarrow \infty$. It is conveninent to work
in momentum space by defining fourier components for any
bilocal field $A(\vm_1, \vm_2)$ by
\eqn\twenty{A(\vm_1, \vm_2) = \sum_{\vn_1,\vn_2}
\tA (\vn_1,\vn_2)~e^{{2\pi i \over M}(\vn_1\cdot \vm_1 +
\vn_2 \cdot \vm_2)}}
Then the action $\cS$ becomes
\eqn\twoone{\eqalign{
\cS = N & M^d[\sum_\vn (p_n^2 + m_0^2)\tsigma(\vn, -\vn)
+ {\lambda_0 \over 2}\sum_{\vn_1,\vn_2,\vn_3}
\tsigma(\vn_1,\vn_2)\tsigma(\vn_3,-(\vn_1+\vn_2+\vn_3))] \cr
& -\half (N - 2 M^d)~{\rm Tr} \log~\sigma}}
where $m_0$ and $\lambda_0$ are the dimensionless mass and coupling
constant respectively (with appropriate powers of the cutoff
multiplying the dimensional quantities $m$ and $\lambda$) and
\eqn\twoonea{p_n^2 = 4 \sin^2 {n \over 2}~~~~~n = |\vn|}
The saddle point is translationally invariant so that 
\eqn\twotwo{\tsigma (\vn_1,\vn_2) = \xi (\vn_1) \delta_{\vn_1,-\vn_2}}
With this ansatz, the term in the action becomes
\eqn\twothree{\cS = N[ M^d \sum_\vn (p_n^2 + m_0^2)  \xi (\vn)
+ M^d  {\lambda_0 \over 2}\sum_{\vn \vn'} \xi (\vn) \xi (\vn ')
-\half \sum_\vn \log~\xi (\vn)] + O(1/N)}
so that the saddle point becomes
\eqn\twofour{ \xi (\vn) = {1\over 2 M^d}
{1\over p_n^2 + m_0^2 + \lambda_0 s}}
where
\eqn\twofive{ s = \sum_\vn \xi (\vn)}
Equation \twofive\ is of course the lowest order (in $1/N$)
propagator $< \vphi (\vx) \cdot \vphi (\vy) >$ in 
momentum space.
The gap $s$ is then determined by a gap equation
\eqn\twosix{ s = {1\over 2 M^d}
\sum_\vn {1\over p_n^2 + m_0^2 + \lambda_0 s}}
In the continuum limit the equation \twosix\ reads
\eqn\twosixa{s = {1\over 2}\int {d^dp \over (2\pi)^d}
{1\over p^2 + m^2 + \lambda s}}
The saddle point value of the action is then given by
\eqn\twoseven{\cS_0 = {N \over 2}[\sum_\vn
\log~(p_n^2 + m_0^2 + \lambda_0 s) - M^d \lambda_0 s^2]}
upto an unimportant constant.

The theory is on the critical surface when the renormalized mass 
vanishes. In this critical theory, one has
\eqn\twosevena{\xi_c (p) = {1 \over 2 |p|^2}}
For any dimension $d$ the point $\lambda_0 =0 $ is of course 
a fixed point. For $d=3$ this gaussian fixed point is unstable
and there is an IR stable fixed point at a finite value of
$\lambda_0$, as will be explained in a following section.

\subsec{Leading ${1\over N}$ correction and propagator}

The ``classical'' action $\cS$ evaluated at the saddle point
already has a $O(1)$ piece which is given by
\eqn\twoeight{\cS_1^{(1)} = - M^d \sum_\vn
\log~(p_n^2 + m_0^2 + \lambda_0 s)}
As we will see below, the extra power of the number of lattice 
points $M^d$ in \twoeight\ is significant, as is its
sign.
Other contributions to this order are obtained by expanding the
collective field as
\eqn\twoeighta{\tsigma(\vn_1,\vn_2) = \tsigma_0(\vn_1,\vn_2)
+ {1\over \sN} \teta (\vn_1,\vn_2)}
Then the quadratic action for $\teta$ is
\eqn\twonine{\eqalign{\delta \cS = {\lambda_0 M^d \over 2}
& \sum_{\vn_1,\vn_2,\vn_3} \teta (\vn_1,\vn_2)
\teta (\vn_3,-(\vn_1+\vn_2+\vn_3))\cr
& + {1\over 4}
\sum_{\vn_1,\vn_2} \xi^{-1} (-\vn_1) \xi^{-1} (\vn_2)
\teta (\vn_1,\vn_2) \teta (-\vn_2, -\vn_1)}}
The continuum expression for the quadratic action is
\eqn\twoninea{\eqalign{ \delta \cS =
{1\over 4} \int  & {d^d p_1 d^d p_2  d^d p_3 d^d p_4 \over (2\pi)^{2d}}~
\{~   {2\lambda \over (2\pi)^d} \delta^{(d)} (\vp_1+\vp_2+\vp_3+\vp_4) \cr
 & + \delta^{(d)}(\vp_3+\vp_2)\delta^{(d)}(\vp_4+\vp_1) \sigma_0^{-1} (\vp_1)
\sigma_0^{-1} (\vp_2) ~\}~ \teta (p_1,p_2)~ \teta (p_3,p_4)}}
where
\eqn\twonineb{\sigma_0 (\vp) = M^d \xi (\vn)~~~~~~\vp = {2\pi \vn
\over Ma}}
From \twoninea\ one can calculate the two point function of the
fluctuations \mello\ 
\eqn\twoninec{ <\teta (p_1,p_2) \teta (p_3,p_4)> 
= \delta^{d}(p_1+p_4)  \delta^{d}(p_2+p_3)\sigma_0 (p_1) \sigma_0
(p_2) - G(p_1,p_2,p_3,p_4)}
where
\eqn\twonined{G(p_1,p_2,p_3,p_4) = 
{2\lambda \sigma_0 (p_1)\sigma_0 (p_2)\sigma_0 (p_3)\sigma_0 (p_4) \over
1 + 2\lambda \int {d^dk \over (2\pi)^d}\sigma_0 (-k)\sigma_0
(k-p_1-p_2)}
~\delta^d(p_1+p_2+p_3+p_4)}

\subsec{$O(1)$ Partition function}

As advertised in the introduction, there is an interesting
cancellation between contributions coming from the $O(1)$ terms in the
``classical'' action and those coming from integrating out the
fluctuations. 

Consider for example the free theory at $\lambda_0 =
0$. Then from the formulation in terms of the fields $\vphi$ the
partition function may be exactly evaluated
\eqn\thirty{ \log~Z = {N \over 2}[\sum_\vn
\log~(p_n^2 + m_0^2)]}
Clearly this exact answer is reproduced by the $O(N)$ classical
value of the collective field action in \twoseven. Therefore the
$O(1)$ contributions from \twoeight\ should cancel whatever one
gets by integrating out the fluctuations $\teta$. This is
straightforward to check. The $O(1)$ contribution to the effective 
action coming from the fluctuations is given by (for $\lambda_0 = 0$)
\eqn\threeone{S_1^{(2)} = {1\over 2}\sum_{\vn_1,\vn_2}
\log~(p_{\vn_1}^2~p_{\vn_2}^2) = M^d \sum_\vn \log~p_\vn^2}
Adding the contribution $S_1^{(1)}$ from \twoeight\ (with
$m_0=\lambda_0 = 0$ ) we see that the total $O(1)$
contribution to the partition function is
\eqn\threetwo{(\log~Z)_1 = -(S_1^{(1)} + S_1^{(2)}) = 0}
as expected.  From the collective field theory point of view, at any
stage of the $(1/N)$ expansion, there are two contributions, one from
the ``classical'' action and one from the fluctuations.  For the free
theory these should cancel precisely.

For $\lambda_0 \neq 0$ the situation is more complicated. Here there
are nonzero subleading terms in the partition function. Now
contributions from the classical action {\it partially} cancel those
coming from the fluctuations. Significantly, in the continuum limit
the ultraviolet divergent terms cancel at the level of leading $1/N$
correction - this is evident from the fact that at this level the
effect of a nonzero $\lambda_0$ is to simply change the mass gap, and
this does not affect the ultraviolet behavior. These cancellations
have important consequences for a holographic interpretation.

\subsec{Nature of interactions}

The cubic and higher order interactions in the collective field theory
come entirely from the Jacobian factor. On the finite lattice this has
the structure
\eqn\tone{(N-2M^d){\rm Tr}[\log ~\sigma_0 - \sum_{k=2}^\infty
{(-1)^k \over k N^{k/2}}(\sigma_0^{-1}\eta)^k]}
In each order of ${1\over {\sqrt{N}}}$ there are generically two terms
which come from the two terms in the overall coefficient.

The interactions have an interesting scale invariant form in the
critical theory. In this case the classical value of the collective
field $\sigma_0$ is simply the massless propagator and it is
straightforward to see that the term which contains $k$ factors of the
fluctuation $\eta$ has the following form in the continuum and
thermodynamic limits
\eqn\fone{\int  \prod_{i=1}^k (d^d x_i)  
[\partial_{x_1}\partial_{x_2} \eta
(x_1,x_2)\partial_{x_2}\partial_{x_3} 
\eta (x_2,x_3)
\cdots \partial_{x_k} \partial_{x_1} \eta (x_k, x_1)]}
It is interesting to note that the cubic and higher order interactions
do not depend on $\lambda$.
The coupling constant of the collective field theory 
is ${1\over {\sqrt{N}}}$ as expected.

\subsec{Fixed points for $d=3$}

For the special case of $d=3$ there is a nontrivial fixed point 
away from $\lambda = 0$. This may be seen in the collective theory
in the following way.

To arrive at the IR fixed point one has to first put the theory on the
critical surface by tuning the renormalized mass to zero. In this case
the saddle point value of the bilocal field is, in momentum space
\eqn\fone{\sigma_0(p) = {1\over 2p^2}}
The momentum space propagator for the bilocal field fluctuation now
reads
\eqn\ftwo{\eqalign{<\teta (p_1,p_2) \teta (p_3,p_4)> 
= {1\over p_1^2 p_2^2}[ \delta^{d}(p_1+p_4) &  \delta^{d}(p_2+p_3) \cr
& -{2\lambda \over p_3^2 p_4^2}{\delta^d (p_1+p_2+p_3+p_4)\over
1 + 2\lambda \int {d^dk \over (2\pi)^d}{1\over (p_1+p_2-k)^2 k^2}}]}}
It is clear from the form of the classical solution that there is no
anomalous dimension for the fundamental field $\phi^i$, whereas the
expression \ftwo\ shows that there would be anomalous dimensions for
composite operators in general. The first term in \ftwo\ is the
contribution of free field theory. Thus the second term may be used to
define a dimensional {\it running} coupling constant
\eqn\fthree{\alpha (p) = {\lambda \over
1 + 2\lambda \int {d^dk \over (2\pi)^d}{1\over (p-k)^2 k^2}}}
The basic integral is given by
\eqn\ffour{ I(p) =  \int {d^dk \over (2\pi)^d}{1\over (p-k)^2 k^2}
= {1\over 8 |p|}}
So that
\eqn\ffour{\alpha (p) = {\lambda \over 1 + {\lambda \over 4 |p|}}}
Clearly in the infrared
\eqn\ffive{ {\rm Lim}_{p \rightarrow 0} \alpha (p) = 4 |p|}
the running coupling becomes independent of the original bare coupling
of the theory. 
The {\it dimensionless} running coupling $\alpha (p)/p$
approaches a constant nemerical value, 4. Alternatively one may define
a dimensionless renormalized coupling at some scale $\mu$, $g(\mu)$ by the
relation
\eqn\fsix{\mu g(\mu) = {\lambda_0 \Lambda \over
1 + {\lambda_0 \Lambda \over 4 \mu}}}
where $\Lambda$ denotes the momentum space cutoff of the bare theory
and $\lambda_0$ the bare dimensionless coupling. Then as one
approaches the continuum limit $\Lambda \rightarrow \infty$ the
renormalized coupling $g$ tends to a fixed value $4$ independently of
$\lambda_0$. Thus $g = 4$ is an infrared fixed point.
The full beta function, at leading order of large-N expansion can be
read off from \fsix\
\eqn\fsixa{\beta (g) = \mu {\partial g \over \partial \mu}
= -g ( 1 - {1\over 4} g)}

The scaling behavior of the correlators \ftwo\ at this IR fixed point
may be read off by considering the continuum limit approached by
starting with any arbitrary bare coupling $\lambda_0$. Instead of
considering the bilocal field by itself it is instructive to consider
various moments which define local composite operators as in \sevenc.
Consider the simplest such operator which is the scalar composite
\eqn\fseven{\sigma(\vx,\vx) = \vphi (\vx) \cdot \vphi (\vx)}
The foruier components of this composite may be expressed in terms of
the fourier components $\tsigma (p,q)$ as
\eqn\feight{\tsigma (k) = \int [d^3 x] e^{i \vk \cdot
\vx}\sigma(\vx,\vx)
= \int {d^3 q \over (2\pi)^3} \tsigma (k-q,q)}
Thus the connected two point function of the composite operator is given by
\eqn\fnine{<\tzeta (k_1) \tzeta (k_2)> = \int {d^3 q \over (2\pi)^3}
{d^3 q' \over (2\pi)^3}<\teta (k_1-q,q)\teta (k_2-q',q')>}
where $\tzeta (k) $ denotes the fluctuation in $\tsigma(k)$. Using 
\ftwo\ and \ffour\ this may be easily evaluated to yield
\eqn\ften{ <\tzeta (k_1) \tzeta (k_2)> = {\delta (k_1+k_2) \over
8|k_1|} [1 - {1\over 8|k_1|}{ 2\lambda \over 1 + {2\lambda \over
8|k_1|}}] =  
{\delta (k_1+k_2) \over \lambda + 8|k_1|}}
At the trivial fixed point one gets
\eqn\feleven{<\tzeta (k_1) \tzeta (k_2)> = 
{\delta (k_1+k_2) \over 8|k_1|}}
which implies that the dimension of the operator $\sigma (\vx,\vx)$ is
$1$. To see the behavior at the nontrivial fixed point one has to
rewrite this expression in terms of the bare dimensionless coupling
$\lambda_0$ and the momentum cutoff $\Lambda$ and perform the
continuum limit with nonzero $\lambda_0$. This yields
\eqn\ftwelve{<\tzeta (k_1) \tzeta (k_2)> = \delta (k_1+k_2)
[{1 \over \lambda_0 \Lambda}
- {8 |k_1| \over \lambda_0 \Lambda^2} + O({|k_1|^2 \over \Lambda^3})]}
In the limit $\Lambda \rightarrow \infty$ the first term gives rise to
a short distance contact term which has to be subtracted in the
renormalized theory. The nontrivial part scales as $|k_1|$ which means
that the dimension of the operator is $2$ at the nontrivial fixed point.

\newsec{Holographic correspondence}

Our proposal is that the collective field theory described above
provides a description of the singlet sector of the vector
model in terms of a $(d+1)$ dimensional theory of higher spins.
The basic idea is to write the collective field as a function of the
center of mass coordinates $\vu$ and the relative coordinate $\vv$ as
in \six\ and \seven. One can then expand the field $\sigma$ as
\eqn\threethree{\sigma(\vu, \vv) = \sum_{l,m}
\sigma_{l\vm} (\vu,r)~Y_{l\vm} (\theta_i)}
where we have written the $d$ relative coordinates $\vv$ in 
terms of its magntitude $r$ and $(d-1)$ angles 
$\theta_1 \cdots \theta_{d-1}$. $Y_{l\vm}(\theta_i)$ denote the
spherical harmonics on $S^{d-1}$. Since
the original field $\sigma (\vx,\vy)$ is symmetric under interchange of
$\vx$ and $\vy$, it should be symmetric under $\vv \rightarrow -\vv$.
This means that only even (or zero ) values of $l$ appear in the
expansion \threethree. Thus the collective field is equivalent to
a collection of higher spin fields living in $d+1$ dimensions spanned
by $(\vu,r)$ and there is exactly one field for each even spin. 

Note that if we had a $U(N)$ rather than a $O(N)$ symmetry one would
have odd spins as well.

For $d=3$ we thus have a four dimensional theory. When the vector
model is at one of its fixed points, the theory is conformally 
invariant and has a symmetry group $SO(4,1)$. It is then natural
to expect that the four dimensional theory is defined on $AdS_4$
which has the same isometry. We will see later in what sense this
is true.

In the remaining part of this section we will discuss several issues
which points towards an interpretation of the collective field
theory for the fixed point models 
as a {\it holographic} theory defined on the boundary of $AdS_4$.

\subsec{Finite temperature thermodynamics}

One of the crucial aspects of holography is that the high temperature
thermodynamics of the bulk theory in $(d+1)$ dimensions is appropriate
to a theory in $d$ dimensions. Furthermore, the result involves $N$
which is the coupling constant of the bulk theory. This leading result
cannot come from counting of the states of the bulk theory since $N$
appears in the latter only through the coupling constant and in a
${1\over N}$ expansion one would expect a $O(1)$ answer which reflects
that the propagating modes live in $(d+1)$ dimensions.
In known examples of holography, however, a $N$ dependent answer
characteristic of a $d$ dimensional theory comes from the fact that
the bulk theory is typically a theory of gravity and its high
temperatrure properties are dominated by black holes whose entropies
is proportional to their areas and whose thermodynamics is appropriate
to that of theory in $d$ dimensions. Furthermore the black hole
entropy is a ``classical'' effect and 
goes as the inverse of the square of the coupling constant and
therefore contains the right power of $N$.

In the previous section we have calculated the leading order and
the $(1/N)$ corrections to the partition function of the collective
field theory defined on a periodic lattice with $M$ sites in each direction.
We now use these results to discuss the finite temperature behavior.
To do this all we have to do take $M$ large but consider different
lattice spacings in the ``space'' and the ``euclidean time'' directions.
Finally we have to consider a continuum limit and a thermodynamic
limit in which the physical extent of the euclidean time direction is
a finite quantity $\beta = 1/T$ 
while those in the space directions are $L$ with $L \gg \beta$. 

First consider the gaussian fixed point at $\lambda = 0$ in any
number of dimensions.
The finite temperature free energy may be read off from \twoseven\ in
a standard fashion. The expression \twoseven\ has a leading divergent
term which is extensive, proportional to $L^{d-1}\beta$ - the
coefficient being the ground state energy density. The next subleading
term, which we denote by $\cS'$, 
is proportional to $L^{d-1}$ and has the form
\eqn\threefive{\cS' = - N L^{d-1} \int {d^{d-1} p \over 
(2\pi)^{d-1}}\log~(1 - e^{-\beta |p|})}
This is then related to the thermodynamic free energy $F$ by
\eqn\threesix{ F = {1\over \beta} \cS'}
It is clear that 
\eqn\threeseven{ F \sim N L^{d-1} T^{d}}
which is nothing but the free energy of $N$ species of massless
particles in $d-1$ space dimensions. In particular the entropy scales as
\eqn\threeeight{ S \sim N L^{d-1} T^{d-1}}

From the point of view of the vector model this result is of course
obvious. However, from the point of view of the collective field
theory this is a rather nontrivial result. As we saw, we can interpret
this theory as a theory of higher spin fields in $d+1$ dimensions.
Naively one would expect that the thermodynamics would be the one
appropriate to $d$, rather than $(d-1)$, space dimensions and should
be of $O(1)$ as explained above. 

The point is that
all these expectations are based on the usual situation where the
leading thermodynamic free energy comes from the ``one loop'' 
contribution. In the present context this is the leading ${1\over N}$
correction, which actually gives a $O(1)$ contribution to the free
energy. What we saw above, however, is that there is a {\it classical}
contribution to the free energy which is proportional to $N$ 
{\it which leads not only to a nontrivial internal energy, but a
nontrivial entropy proportional to } $N$. 
In fact
there are no ${1\over N}$ corrections to this result for the $\lambda = 0$
theory due to the complete cancellation discussed above. For $\lambda
\neq 0$ the cancellation is not complete and there are finite $O(1)$
corrections. However the divergent terms which contributes to the
vacuum energy cancels. The finite tempertaure behavior of the
interacting $O(N)$ model was discussed long time ago in
\ref\dolan{L. Dolan and R. Jackiw, {\it Phys. Rev.} {\bf D 9} 3320}
and more recently in \ref\sachdev{S. Sachdev, {\it 
Phys.Lett.} {\bf B309} 285.}.

The fact that the leading thermodynamics comes from the classical
contribution to the action is reminiscent of the Gibbons-Hawking
calculation of the entropy of a black hole.  As will be seen below,
the space-time interpretation of the four dimensional collective field
theory does not appear to be straightforward and it is difficult to
identify what kind of space-time configurations give rise to this
classical contribution. Nevertheless our result strongly suggests that
the four dimensional bulk theory of higher spin fields have black
holes.

\subsec{Conformal transformations and $AdS$}

While the fields $\sigma_{lm}(\vu,r)$ do represent higher spin fields
in $d+1$ dimensions, they are not the standard higher spin fields as
discussed in \higher, but related by some field redfinition.  This may
be seen from the fact that the quadratic action is not diagonal in the
spins. A spin $l$ field mixes with spins $l \pm 2$. We do not know
what is the exact field redefinition which relates these components to
the standard fields of higher spin theories.

An important indication of this fact comes from an examination of the
transformation properties of the collective field under conformal
transformations on the boundary. Consider for example the theory at
the gaussian fixed point. 
From the known conformal transformations it follows that
the transformations of the bilocal field $\sigma (\vx,\vy)$ are
given by
\eqn\conone{\eqalign{& \delta_D \sigma = 
-\alpha(x^i {\partial \over \partial x^i} + y^i {\partial \over \partial y^i}
+ \Delta)\sigma (\vx,\vy) \cr
& \delta_T \sigma = t^i({\partial \over \partial x^i} +
{\partial \over \partial y^i})\sigma \cr
& \delta_R \sigma = \theta^{ij}(x^i {\partial \over \partial x^j}
- x^j {\partial \over \partial x^i} + y^i {\partial \over \partial y^j}
- y^j {\partial \over \partial y^i})\sigma \cr
& \delta_S \sigma = \{ [2(\epsilon \cdot x) x^i - |x|^2 \epsilon^i]
{\partial \over \partial x^i} +
[2(\epsilon \cdot y) y^i - |y|^2 \epsilon^i]
{\partial \over \partial y^i} + \Delta \epsilon \cdot (x + y) \}\sigma}}
Here $\Delta$ is the scaling dimension of $\sigma$ and $\alpha, t^i,
\theta^{ij}$ and $\epsilon^i$ are the parameters of dilatations,
translations, rotations and special conformal transformations respectively.
Rewriting these expressions in terms of $\vu$ and $\vv$ we get
\eqn\contwo{\eqalign{
\delta_D \sigma = & -(u^i {\partial \over \partial u^i} + r
{\partial \over \partial r} + \Delta)\sigma \cr
 \delta_T \sigma = & t^i {\partial \over \partial u^i} \sigma \cr
 \delta_R \sigma = & \theta^{ij}(u^i {\partial \over \partial u^j}
- u^j {\partial \over \partial u^i} +L_{ij})\sigma}}
\eqn\contwol{\eqalign{
 \delta_S \sigma =  \{ [2(\epsilon \cdot u) u^i - |u|^2 \epsilon^i
-r^2 \epsilon^i]{\partial \over \partial u^i}  
+ &  2 (\epsilon \cdot u) r {\partial \over \partial r} 
+ 2 \Delta (\epsilon \cdot u)
 + (\epsilon^j u^i - u^j \epsilon^i) L_{ji} \}\sigma \cr 
&  + 2 (\epsilon \cdot v) v^i {\partial \over \partial u^i} \sigma}}
where
\eqn\confour{L_{ij} =  v^i{\partial \over \partial v^j} -
v^j {\partial \over \partial v^i}}
To see the action on the individual components $\sigma_{lm} (\vu,r)$
one needs to substitute these expressions in the expansion
\threethree.
It is clear from the above expressions that the dilatations, translations
and rotations acts on the components $\sigma_{lm} (\vu, r)$ diagonally,
i.e. the action does not mix up various spins. The factor of $L_{ij}$
in the transformation $\delta_R \sigma$ mixes fields of different $m$
for the {\it same} $l$ exactly as rotation generators should. On the
other hand, the last term in the 
special conformal transformation on $\sigma$ shows
that this mixes up fields with {\it different} spin.

In fact, if we define new component fields
\eqn\conthree{ \chi_{lm}(\vu,r) = r^{l + \Delta}\sigma_{lm} (u,r)}
the generators for dilatations, translations and rotations on
$\chi_{lm}$ are {\it exactly} the generators of the corresponding
isometries on tensor fields of rank $l$ defined on a $AdS$ space, with
the metric given in \eight,
\eqn\confoura{ds^2 = {1\over r^2}[dr^2 + d\vu \cdot d\vu]}
In particular this means that the magnitude of the relative coordinate
behaves as a scale, as it should.

For special conformal transformations, the story is different. Here
all the terms in \contwol, {\it except the last
term} are the correct expressions for generators of the corresponding
isometries of the metric \confoura. The last term, however, clearly
mixes different spins.

The correct higher spin fields \higher,\higherb,\higherc,\higherd,
however, transform homogeneously under the Killing isometries of $AdS$
and do not mix up fields with different spin. This shows that the
fields $\sigma_{lm}(\vu,r)$, while containing the complete physics of
higher spin theories are not themselves the correct higher spin
fields. 

In fact there are indications that the correct higher spin fields are
related to the components $\sigma_{lm}(\vu,r)$ by nonlocal
transformations. This may be seen from various points of view. The
exercise we have done above is in fact an attempt to rewrite conformal
transformations on a pair of vectors $(\vx, \vy)$ as isometries in a
$AdS$ space by identifying the correct coordinates in the latter. It
is straightforward to see that this works for dilatations,
translations and rotations with the identification of $\vu$ and $r$ 
as the coordinates in $AdS$ as in the metric \confoura. However this
cannot work for the special conformal transformation. To see this
consider the case of $d=1$. In this case 
collective field should contain only one field in $AdS$ since there is
no spin in $AdS_2$. The special conformal transformations are then
\eqn\conaone{\delta x = \epsilon x^2 ~~~~~~~\delta y = \epsilon y^2}
which leads to
\eqn\conatwo{\delta u = \epsilon(u^2+v^2)~~~~~\delta v = 2\epsilon uv}
This is to be compared with the corresponding Killing isometry of the
metric \confoura\ for $d=1$, viz.
\eqn\conathree{\delta' u = \epsilon( u^2 - v^2)~~~~~\delta v = 2\epsilon uv}

Another indication comes from the relationship of the components
$\sigma_{lm}$ with the infinite set of conserved currents in the
vector model (at $\lambda = 0$) \ref\anselmi{D. Anselmi,
{\it Class.Quant.Grav.} {\bf 17} (2000) 1383, {\tt hep-th/9906167}.}
\higherd. These currents are symmetric and traceless and given by (in $d=3$)
\eqn\conafour{J_{i_1 \cdots i_s}
= \sum_{k=0}^s {(-1)^k (\partial_{i_1}\cdots \partial_{i_k}\phi)
(\partial_{i_{k+1}}\cdots \partial_{i_s}\phi)
\over \Gamma (k+1) \Gamma (k +{1\over 2}) \Gamma (s-k+1)
\Gamma (s-k + {1\over 2})} - {\rm traces}}
These currents are conserved
\eqn\conafive{\partial^{i_1}J_{i_1 \cdots i_s} = 0}
These currents can be expressed in terms of the collective
field. Consider for example the first few currents. These may be
expressed as follows
\eqn\conasix{\eqalign{J_0 = & \sigma (u,v)_{v=0} \cr
J_{ij} = & [{\partial^2 \over \partial v^i \partial v^j}
- {1\over 3} \delta^{ij} {\partial^2 \over \partial v^k \partial v_k}]
\sigma(\vu,\vv)|_{v=0} - {1\over 2}[{\partial^2 \over \partial u^i \partial u^j}
- {1\over 3} \delta^{ij} {\partial^2 \over \partial u^k \partial u_k}]
\sigma(u,0)}}
These currents transform homogeneously under conformal transformations.

The collective field expansion in \threethree\ can be also reorganized
in terms of derivatives of the form which appear in \conasix, since
spherical harmonics are in one-to-one correspondence with traceless
symmetric tensors made out of products of $v^i$. Thus we have
expansions of the form
\eqn\conaseven{\eqalign{\sigma(\vu,\vv) = & [\sigma (\vu,0) + {1\over 6}r^2
({\partial^2 \sigma \over \partial v^2})_{v=0} + O(r^4)] \cr
& + {1\over 2}r^2 (\hv^i \hv^j - {1\over 3} \delta^{ij})[
({\partial^2 \sigma \over \partial v^i \partial v^j})_{v=0} 
- {1\over 3} \delta^{ij} 
({\partial^2 \sigma \over \partial v^2})_{v=0} + O(r^2)]}}
where we have performed a Taylor expansion in $v^i$ and reorganized it
in terms of traceless symmetric products of the unit vectors $\hv^i$.
Thus the components $\sigma_{l,m}$ are given by
\eqn\conaeight{\eqalign{ \sigma_{00} & \sim 
[\sigma (\vu,0) + {1\over 6}r^2
({\partial^2 \sigma \over \partial v^2})_{v=0} + O(r^4)]\cr
\sigma_{1m} & \sim [
({\partial^2 \sigma \over \partial v^i \partial v^j})_{v=0} 
- {1\over 3} \delta^{ij} 
({\partial^2 \sigma \over \partial v^2})_{v=0} + O(r^2)]}}
Comparing \conaeight\ and \conasix\ it is clear that the fields 
$\sigma_{lm}(u,0)$ {\it do not} reduce to the currents $J_{i_1 \cdots
i_l}$. This is the basic reason why these components do not transform
properly under special conformal transformations.

It must be emphasized that the collective field theory contains all
the information contained in the vector model singlet correlators and
hence serves as a {\it complete definition} of the higher spin
theory, including all interactions. However the relationship between
the components of the collective field and the higher spin fields
which propagate independently at $N=\infty$ appears to be rather
nontrivial. The key to uncover the precise relationship is conformal
invariance. We hope to report results about this connection soon
\ref\dejeve{S.R. Das and A. Jevicki, {\it Work in progress}}.

\subsec{Interactions and the bulk theory}

As shown in the previous section, for the critical theory 
the interactions of the collective
field theory are characterized by a coupling constant which is
${1\over {\sqrt{N}}}$ and independent of the bare coupling $\lambda_0$
of the underlying vector model. We now make several comments about how
this may come about in a bulk theory defined on $AdS$ space.

Since the bulk theory contains gravity, it is characterized by a
Newton's gravitational constant $G$ which has dimensions of
$(length)^2$ in four dimensions. In flat space, the interaction terms
in the theory have coefficients which depend on the coupling constant
${\sqrt{G}}$ and the terms have a number of derivatives which make the
action dimensionless. Typically, again for four dimensions, each
${\sqrt{G}}$ is accompanied by a single derivative. In $AdS$ space
however there is another length scale, $R$ where ${1\over R^2}$ is the
constant curvature. Consequently, instead of derivatives there could
be inverse powers of $R$ which account for the correct
dimensions. This is familiar in supergravity in e.g. $AdS_5 \times
S^5$. Here, there is a class of couplings which do not depend on $G$
and $R$ individually, but only on the dimensionless combination ${G
\over R^3}$. By virtue of the $AdS/CFT$ correspondence one has ${G
\over R^3} \sim {1\over N^2}$ ,where the four dimensional dual Super
Yang Mills theory has a gauge group $SU(N)$. Since the gauge theory
coupling $g_{YM}$, is related to the bulk parameters by the relation
$g_{YM} = G^{1\over 4}R^{5\over 4}l_s^{-2}$, such bulk couplings are
computed in terms of gauge theory three point functions which are
completely independent of $g_{YM}$ and only depends on $N$. This is
possible since one is computing three point functions of composite
operators which have nonzero values in free field theory.Indeed in
this particular case, the underlying supersymmetry ensures that the
three point functions of a class of operators are given exactly by
their free field values \ref\adscoupling{
D. Freedman, S. D. Mathur, A. Matusis and L. Rastelli, {\it
Nucl.Phys.} {\bf B546} (1999) 96, {\tt hep-th/9804058};
H. Liu and A. Tseytlin, {\it Nucl.Phys.} {\bf B533} 88,
{\tt hep-th/9804083};
G. Chalmers,
H. Nastase, K. Schalm, and R. Siebelink, {\it Nucl.Phys.} {\bf B540}
(1999) 247, {\tt hep-th/9805105}; S. Lee, S.  Minwalla, M. Rangamani
and N. Seiberg, {\it Adv.Theor.Math.Phys.} {\bf 2} (1998) 697, {\tt
hep-th/9806074.}}.

For the vector model, the conformal field theory is at a fixed point
rather than on a line of fixed points, so that there is no analog of a
gauge theory coupling constant. This is the reason why the couplings
in the collective field theory are characterized only by $N$ and by no
other parameter. The value of the bare coupling drops out since in the
continuum limits one approaches the infrared fixed point.

This fact therefore implies that if the dual theory is a higher spin
theory in $AdS$, then {\it all} the couplings of that theory are
characterized by the dimensionless combination ${G \over R^2}$.
This is a rather nontrivial prediction for the higher spin theory.

\subsec{Propagating modes and Hamiltonian Collective theory}

We have so far considered the euclidean version of the collective field
theory as derived in \jevsak. For the three dimensional vector model
this is a collection of higher spin fields in four dimensions - one
field for each even spin. A component field $\sigma_{lm} (\vu,r)$ has
$2l+1$ components. However, if all the four dimensional fields of the
dual theory are massless - as conjectured - there are precisely two
propagating polarizations for each spin. $\sigma_{lm}$ clearly
contains too many independent propagating modes.

The key reason behind this overcounting is the fact that the euclidean
collective field is a way to organize an infinite set of higher spin
currents in the boundary theory, as indicated above. These are
symmetric and traceless in the three dimensional indices, with the
spin-l current containing $l$ indices, leading to $(2l+1)$ components.
However these currents are {\it conserved} in the $\lambda = 0$
theory, and conserved to leading order in $1/N$ in the interacting
theory,
so that there are $2(l-1)+1$ conditions relating the
components. Thus the number of independent components is $(2l+1)-
(2(l-1)+1) = 2$, which is the correct value for the number of
propagating modes for each spin. This counting can be easily seen to
work in any number of dimensions.

The meaning of all this is that the euclidean collective field theory
must have a gauge invariance which follows from the current
conservation conditions in the vector model. We do not know how to
display this symmetry, but we know it is there because of the
one-to-one correspondence between the spin components of the
collective field and the currents.

This situation is not new and has been encountered before in other
examples of the $AdS/CFT$ correspondence. Consider the cases where the
bulk theory on $AdS_{d+1}$ contains a massless graviton. This has
${(d+1)(d-2) \over 2}$ propagating components. The operator which is
dual to the graviton is the energy momentum tensor of the boundary
theory which has ${d^2 + d-2 \over 2}$ components because of the
tracelessness. However the energy-momentum tensor is conserved so that
the number of independent polarizations is ${d^2 + d-2 \over 2}-d
= {(d+1)(d-2) \over 2}$. In the coordinates of \confoura, the CFT is
defined on the boundary at $r=0$. The energy momentum tensor then
computes correlators of the graviton field $h_{\mu\nu}$ in a gauge
where $h_{r\mu} = 0$, but this gauge still retains some gauge
symmetries. 

Our situation is rather similar. In fact, given the conserved
currents in the vector model one may construct bulk fields using a
bulk-to-boundary propagator. As shown in \higherd, the conservation of
currents then lead to gauge conditions on the bulk fields.

It is not surprising to find that the euclidean collective field
contains redundant degrees of freedom. The collective field
theory we have considered reproduces all singlet correlators of the
theory. However among these correlators are those which receive
contributions from {\it non-singlet} intermediate states. The simplest
example is $< \sigma (x,y) >$ itself, which is the propagator of the
elementary field $\vphi$. On the other physical propagating states in
the bulk must be singlet states. 

To look at the propagating modes it is instructive to consider the
hamiltonian version of collective field theory \jevsaka,
\ref\jevrod{A. Jevicki and J. Rodrigues,
{\it Nucl.Phys.}{\bf B230} (1984) 317.}. In this formulation the 
collective fields are Schrodinger picture operators $\psi (\tx,\ty)$ 
defined by 
\eqn\yone{\psi (\tx, \ty) = {1\over N} \vphi (\tx) \cdot \vphi (\ty)}
and their canonically conjugate momenta $\Pi (\tx,\ty)$.
Here $\tx,\ty$ denote the {\it spatial} (i.e. $(d-1)$ dimensional)
components of the space time locations. These operators create all the
singlet states of the theory, whose dynamics is governed by the
collective field Hamiltonian.
\eqn\yten{H = 2\Tr (\Pi \psi \Pi) + V_{coll}} 
with
\eqn\yeleven{V_{coll} = {1\over 2}\int dx [- \nabla_x^2 
\psi (\tx , \ty)|_{\tx=\ty} +m^2 \psi (\tx,\tx) +{\lambda \over 2}(\psi
(\tx,\tx))^2]
+ {1\over 8} \Tr \psi^{-1}}
and
\eqn\ytwelve{\Pi (\tx, \ty)= {\delta\over \delta \psi
(\tx , \ty )}}
being the canonically conjugate variable. In \yten\-\ytwelve\ $\psi$
should be regarded as a matrix in $\tx,\ty$ and the trace refers to
the trace of this matrix.

One may expand the corresponding Heisenberg 
picture operator in a
manner similar to the spherical harmonic exapnsion of the euclidean
collective field
\eqn\ytwo{\psi (\tx,\ty;t) = \sum_{l,m} \psi_{lm}(\tu,r,t) Y_{lm}}
where as usual 
\eqn\ythree{\tu = {1\over 2} (\tx + \ty)~~~~\tv = {1\over 2} (\tx -
\ty)~~~~r^2 = \tv \cdot \tv}
In \ytwo\ $Y_{lm}$ are spherical harmonics on a $S^{d-2}$ rather than
on $S^{d-1}$. Consequently the number of components of $\psi_{lm}$ for
a given $l$ is exactly the same as the number of propagating
polarizations of a massless spin-$l$ field in $d+1$ dimensions. For
example for $d=3$ for a given $l$ we have precsiely two values of $m$,
i.e. $m = \pm l$ which count the two polarizations of massless four
dimensional fields with any spin.
This hamiltonian collective field theory therefore correctly counts
the propagating modes of higher spin fields. 

The manner in which the action formulation reduces to the canonical
,hamiltonian representation is interesting and rather nontrivial. The
Langragian formulation was characterized by being bi-local in time as
in space while the canonical, Hamiltonian formulation is local in
time.  The reduction from one to another
\eqn\yfourteen{
\sigma (\tx, t ; \tx' t' ) \longrightarrow 
\psi (\tx , \tx' ; t )}
involves a formal reduction in the number of degrees of freedom as we
have seen.One has indications that this reduction can be understood in
terms of a gauge principle in analogy with a connection between a
covariant and canonical gauge description of gravity.  We should also
emphasize implications on thermodynamics contained in the two
descriptions. By its nature the canonical description naturally leads
to a thermodynamics with entropy of order one rather than of order $N$
and would seemingly miss one of the main ingredients of
holography. (There is a vacuum energy of $O(N)$ but not an entropy).
The reason behind this may be gleaned from an
understanding of the free theory. Here the exact entropy is of order
$N$ and clearly counts the number of states created by the elementary
fields $\vphi$. In other words this leading classical contribution to
the thermodynamics comes from the {\it non-singlet} states of the
theory. Analogous states have an interpretation of winding modes in
matrix theories and they are not contained in the hamiltonian
collective theory. On the other hand as we have seen the euclidean
collective field theory does capture the contribution from these
non-singlet states.  The importance of nonsinglet states for
thermodynamics also makes its appearance in the $d=1$ matrix model
\kkk.

We therefore see that the nonsinglet states of the vector model
correspond to non-propagating modes in the bulk. The thermodynamics in
the bulk description comes from a classical contribution in the
euclidean collective field theory and hence from these non-propagating
modes. This is consistent with the conjecture that in this model, like
in other string theory examples, the thermodynamics is dominated by
black holes, which are of course examples of condensates of
nonpropagating modes. The relationship between nonsinglet states,
nonpropagating modes, black holes etc. has been a matter of
considerable discussion in matrix models. For vector models, the
availabilty of a tractable euclidean collective theory provides an
opportunity to understand this important issue. 

\newsec{Conclusions}

We have argued that the {\it euclidean} bilocal collective field is
capable of describing a higher spin theory of a single field for each
even spin in one higher dimension. However while the spherical
harmonic decomposition of the bilocal field gives the correct count of
the higher spins, these are {\it not} the standard higher spin fields.
We suspect that there is a possibly nonlocal field redefinition
between $\sigma_{lm}$ and the standard fields.

We have also argued that the euclidean collective field contains more
degrees of freedom than the {\it propagating} modes. On the other hand
the Hamiltonian collective theory precisely counts the propagating
modes. The distinction is important, since as we have found, the
leading thermodynamics in fact receives contributions from nonsinglet
states. In the bulk description this means that nonpropagating
backgrounds dominate the thermodynamics (which is why the result is
{\it classical}). One of course knows examples of this in string
theory realizations of the $AdS/CFT$ duality : here black holes
provide the right thermodynamics.

One of the most interesting aspects of the duality between vector
models and higher spin theories is that one has a fully consistent
quantum theory in four dimensions which contain gravity - and this is
not a string theory. In string theory, however, higher spin fields are
massive with the masses determined by the string length. Thus at
distances much larger than the string length, string theory is
essentially (super)-gravity. In this case, there is no analog of
a string length. This means that there is no separation
of the spin-2 field with all the other higher spin fields. In other
words, considered as a theory of gravity there is no reason to expect
that this theory has nice locality properties. 

Nevertheless, it is conceivable that due to unknown reasons there is
a sense in which this theory may be considered as a theory of
gravity \foot{This possibility was suggested by Shiraz
Minwalla}. This may happen as in $AdS_5 \times S^5$ - Yang Mills
duality. Here at the $AdS$ scale, the bulk theory cannot be
{\it a priori} described in terms of a five dimensional theory of
gravity, since the Kaluza-Klein modes from the $S^5$ have the same
scale. However it turns out that one does much better than this naive
expectation. For many purposes, 
the theory can indeed be regarded as five dimensional
gravity even at the $AdS$ scale. This is evidenced by the fact that
the thermodynamics of the Yang-Mills theory is correctly reproduced by
{\it five dimensional} $AdS$-Schwarzschild black holes. We have no
idea whether a similar decoupling holds in this case. 

\newsec{Acknowledgements} S.R.D. would like to thank S. Kachru,
I. Klebanov, S. Mathur,
S. Minwalla, G. Murthy, A. Shapere, 
E. Silverstein, S. Shenker and L. Susskind
for discussions. Some of the results were presented at ``Workshop
on String Theory'' held at Harishchandra Research Institute,
Allahabad, December 12-21, 2002. S.R.D. would like to thank the
organizers and participants of the workshop for providing a
stimulating atmosphere. Thw work of S.R.D. was supported in part by
the U.S. Department of Energy contract DE-FG01-00ER45832. A.J. was
partially supported by the U.S. Department of Energy under contract
DE-FG02-91ER4068.

\listrefs

\end